\documentclass[aps,pra,twocolumn,amsmath,amssymb,revsymb]{revtex4-1}
\usepackage{graphicx}
\pdfoutput=1
\usepackage{color} 
\usepackage{dcolumn}
\usepackage{bm}
\usepackage{braket}
\usepackage[colorlinks=true, urlcolor=blue, citecolor=red]{hyperref}
\usepackage{txfonts}

\definecolor{gray}{rgb}{0.5,0.5,0.5}

\newcommand\mychi{\raisebox{0.35ex}{$\chi$}}

\begin{document}

\title{Towards a heralded eigenstate preserving measurement of
  multi-qubit parity in circuit QED}
\preprint{}

\author{Patrick Huembeli$^{1,2}$}\email[]{patrick.huembeli@icfo.eu}
\author{Simon~E.~Nigg$^{1}$}\email[]{simon.nigg@unibas.ch}
\affiliation{$^{1}$Department of Physics, University of Basel,
  Klingelbergstrasse 82, 4056 Basel, Switzerland}
\affiliation{$^{2}$ICFO - Institut de Ciences Fotoniques,
The Barcelona Institute of Science and Technology,
08860 Castelldefels, Spain}
\date{\today}

\begin{abstract}
Eigenstate-preserving multi-qubit parity measurements lie at the heart of stabilizer quantum
error correction, which is a promising approach to mitigate the
problem of decoherence in quantum computers. In this work we explore a
high-fidelity, eigenstate-preserving parity readout for
superconducting qubits dispersively coupled to a microwave resonator, where the
parity bit is encoded in the amplitude of a coherent state of
the resonator. Detecting photons emitted by the resonator via a current biased
Josephson junction yields information about the parity bit. We
analyse theoretically the measurement back-action in the limit of a
strongly coupled fast detector and show that in general such a parity measurement,
while approximately Quantum Non-Demolition (QND) is not
eigenstate-preserving. To remediate this shortcoming we propose a
simple dynamical decoupling technique during photon detection, which greatly reduces
decoherence within a given parity subspace. Furthermore, by applying a sequence of fast
displacement operations interleaved with the dynamical decoupling
pulses, the natural bias of this binary
detector can be efficiently suppressed. Finally, we introduce the concept of a
heralded parity measurement, where a detector click guarantees
successful multi-qubit parity detection even for finite detection efficiency.
\end{abstract}

\maketitle

\section{Introduction\label{sec:introduction}}
Quantum computers are open quantum
systems: The quantum information carriers -- qubits -- inevitably couple to the outside world and this
coupling leads to decoherence. Quantum error correction (QEC), which aims at correcting the
errors induced by decoherence, is thus necessary for quantum computation. Stabilizer codes~\cite{Gottesman-Thesis} are among the most promising
quantum error correction codes. A common feature of all stabilizer
codes is that errors happening on the physical qubits can be detected
by repeatedly measuring a set of mutually commuting multi-qubit
operators called stabilizer operators. Every detectable error needs to
anti-commute with at least one stabilizer operator. Typically, stabilizer operators
are chosen as elements of the Pauli group, represented by tensor products of single-qubit
operators in the set $\{\openone, X, Y, Z\}$. Here $X$, $Y$ and $Z$
are the three spin-$1/2$ Pauli matrices and $\openone$ is the identity
operator. If in the system under consideration all qubits can be
addressed and controlled individually, then measuring arbitrary multi-qubit Pauli
operators is equivalent, up to single-qubit rotations, to measuring arbitrary tensor products
of operators in the reduced set $\{\openone, Z\}$. The latter task,
which we call parity measurement, is what we focus on in this
work.

To be useful for the purpose of quantum error correction, parity
measurements need to be eigenstate-preserving, e.g. measuring the
parity of the two-qubit state $(1/\sqrt{2})(\ket{ee}+\ket{gg})$ must
not destroy the superposition. Note that this is a stronger requirement than
asking the measurement to be QND, which only requires that repeated
measurements yield always the same
result~\cite{Braginsky-1980}.

Developing multi-qubit parity measurements in superconducting circuits
is a very active area of
research and has been discussed in a number of
previous
works~\cite{Tornberg-2010,Lalumiere-2010,Denhez-2012,Kockum-2012,Nigg-2013,Solgun-2013,Riste-2013,Tornberg-2014,Riste-2015,Corcoles-2015,Govenius-2015,Criger2016,Takita-2016,Blumoff-2016,Cohen-2017-arxiv}. \citet{Blumoff-2016}
succesfully measured the parity of an
arbitrary subset of three superconducting transmon qubits in an approximately eigenstate-preserving
fashion based on the theoretical proposal of~\citet{Nigg-2013}. In this approach in a
first stage, the
parity bit is first mapped onto the {\em phase} of a coherent state of
a microwave field dispersively coupled to the qubits. In a second
stage, the parity bit is mapped onto an ancilla qubit and in a final
third stage, the parity bit is read out by homodyne measurement of the
ancilla qubit.

In the present work we discuss an alternative approach to parity
{\em readout}. This work is motivated by the desire to improve upon two current
limitations of the scheme presented in~\cite{Nigg-2013,Blumoff-2016}, namely the reduction
of parity detection fidelity due to photon leakage and finite ancilla
qubit lifetime. Our proposal can also be seen as an extension
of~\cite{Govia-2015a}, where it was proposed to correlate the parity
of multiple qubits with the {\em amplitude} of a coherent state (either the
vacuum state or a coherent state with finite amplitude) and then detect
the emission of a photon with a microwave photon detector based on a
current biased Josephson junction (CBJJ)~\cite{Romero-2009,Chen-2011,Peropadre-2011,Poudel-2012}. A click of the detector
corresponds to the switching of the CBJJ to
the resistive state. Such an event indicates a certain parity of the
multi-qubit state, while the absence of a photon detection indicates the other
parity with some probability that depends on the measurement time and
the detector efficiency. As presented in
~\cite{Govia-2012,Govia-2014,Govia-2015a}, this elegant scheme however suffers from two important deficiencies. First, while
QND, it leads in general to intra parity-subspace decoherence because
a randomly emitted photon carries with it more information than just
the parity-bit of the multi-qubit state. For the purpose of QEC it
is crucial to limit such information leakage to avoid intra parity-subspace decoherence. 
Second, because one of the two parities is
correlated with a bright state while the other is correlated with the
vacuum in the cavity, the parity detection is
inherently asymmetric: While a click of the photon detector guarantees
the correct parity detection, a no-click event is
ambiguous and the wrong parity can be inferred if the measurement time
is too short or if
the detection efficiency is below unity.

The parity readout proposed in the present work addresses
both of these shortcomings: First, by combining photon detection with
dynamical decoupling, the measurement induced decoherence is reduced. Second, by
periodically swapping the encoding between the two parities and the
dark and bright states of the cavity, the bias of the detector is reduced. With these two modifications, the detection of a photon genuinely
heralds a successful parity detection with minimal back-action induced
intra parity-subspace decoherence. Hence, multi-qubit parity measurements via
direct photon detection become a viable alternative for QEC in an
architecture where fixed frequency qubits are dispersively coupled to a common
bosonic mode.


The manuscript is organized as follows. In Section \ref{sec:Review parity encoding}, we start by reviewing
different methods to encode the parity of multiple qubits into the
state of microwave photons. In Section \ref{sec:review of parity readout}, we briefly
review previous work on how to read out the encoded parity
information. In Section \ref{sec:system and model} we introduce our model for parity readout and make a first qualitative discussion of the measurement back-action in Section \ref{sec:qualtitative discussion}. In Section \ref{sec:Weak coupling limit} we simplify our model. In Section \ref{sec:Characterization of measurement back-action}, the bulk of this work, we analyze the back-action in the
photon detection based parity detection scheme and find that while QND,
the parity measurement is in general not eigenstate preserving,
i.e. it induces intra parity-subspace decoherence. In Section \ref{sec:back-action evasion via DD} we
propose the use of a simple dynamical decoupling scheme to evade the
back-action and quantify the ideal fidelities in this modified scheme
numerically. In Section \ref{sec:detector bias} we propose a way to reduce the detector
bias and show that with these modifications, heralded and unbiased
parity measurement with minimal back-action induced intra
parity-subspace decoherence is feasible. We conclude this section with
an estimate of achievable fidelities for realistic parameter
values. Finally, we conclude with some remarks in
Section \ref{sec:conclusion}.

\section{Review of parity encoding}
\label{sec:Review parity encoding}

The parity operator of $N$ qubits is defined as
\begin{align}\label{eq:5}
\bm P_N=\prod_{n=1}^N\bm\sigma_n^z.
\end{align}
The eigenstates of $\bm\sigma_n^z$ are the computational basis
states. For
a superconducting qubit, they typically correspond to the
two lowest energy eigenstates $\ket{g}$ and $\ket{e}$ so that
$\bm\sigma^z=\ket{e}\bra{e}-\ket{g}\bra{g}$. Sine $(\bm P_N)^2=\openone$, the eigenvalue is either $+1$ or $-1$
and can be interpreted as the parity of the number of
qubits in state $\ket{g}$.  We say that an $N$-qubit state is an even (odd) parity state if it is an
eigenstates of $\bm P_N$ with eigenvalue $+1$ ($-1$).

There exist several methods to encode the parity of a multi-qubit
state into the state of an electromagnetic field. In~\cite{Nigg-2013,
  Blumoff-2016} this was achieved by utilizing the dispersive interaction of
superconducting transmon qubits with the quantized field of a microwave
resonator~\cite{Blais-2004a,Kirchmair-2013a}. This interaction is
characterized by the Hamiltonian term
\begin{align}\label{eq:1}
\bm H_{\rm disp} = \sum_{i=1}^{N} \mychi_i^{}\bm{ \sigma }_i^z \bm{a}^\dag \bm{a},
\end{align}
where $\mychi^{}_i$ is the dispersive frequency shift, while $\bm a$ and $\bm a^{\dagger}$ are the annihilation and
creation operators of the microwave field. The central idea of the
approach of~\cite{Nigg-2013}, is to apply pairs of coherent $\pi$-pulses,
which effectively corresponds to the application of $\bm\sigma_x\equiv \ket{e}\bra{g}+\ket{g}\bra{e}$, to each individual
qubit properly spaced in time such as to control its contribution to
the total phase shift of the cavity, which is initially prepared in a
coherent state with amplitude $\alpha$. Specifically, if the time delay $t_j$
between two $\pi$-pulses on qubit $j$ is chosen as
$t_j=\frac{T}{2}-\frac{\pi}{2\,\mychi_j^{}}$, then under the action of~(\ref{eq:1})
at time $T$ the parity of all
qubits becomes entangled with the phase of the cavity state as
\begin{align}\label{eq:2}
\ket{\phi(T)}=\ket{\alpha_N}\frac{\openone+\bm
  P_N}{2}\ket{\psi_N}+\ket{-\alpha_N}\frac{\openone-\bm P_N}{2}\ket{\psi_N}.
\end{align}
Here $\alpha_N=(-i)^N\alpha$ and $\ket{\psi_N}$ denotes the initial multi-qubit state.
$(\openone\pm \bm P_N)/2$ are the projectors onto the even ($+$) and
odd ($-$) parity subspaces. Selectivity to an arbitrary subset
$\mathcal{S}$ of qubits can be achieved by
instead choosing $t_j=T/2$ for $j\notin\mathcal{S}$~\cite{Nigg-2013}.

In~\cite{Govia-2015a} an alternative method for parity encoding was proposed, which also makes use of
the dispersive interaction~(\ref{eq:1}). Instead of applying control
pulses to the qubits, one applies frequency
multiplexed drives to the cavity initially in the vacuum to
selectively displace the cavity state out of the vacuum conditioned on a
specific parity of the multi-qubit state. This is possible when the
frequency shifts of all the even parity states differ from all the
frequency shifts of the odd parity states but may require slow pulses
to ensure proper frequency selectivity. The encoding thus generated
can be written as
\begin{align}\label{eq:3}
\ket{\varphi(T)}=\ket{0}\frac{\openone+\bm
  P_N}{2}\ket{\psi_N}+\ket{\beta}\frac{\openone-\bm P_N}{2}\ket{\psi_N}.
\end{align}
The amplitude $\beta$ is controlled by the envolop of the applied
drive pulses. Note that the final encodings~(\ref{eq:2}) and~(\ref{eq:3}) are equivalent up to a displacement operation $\bm D(-\beta/2)=\exp[-(\beta/2)\bm
a^{\dagger}+(\beta^*/2)\bm a]$ with $\beta=-2\alpha_N$.

\section{Review of parity readout}
\label{sec:review of parity readout}

To complete the parity measurement, the parity information encoded in
the cavity, as per Eqs.~(\ref{eq:2}) or (\ref{eq:3}), must be read out. We next briefly review two ways to
achieve this. In
Refs.~\cite{Nigg-2013, Blumoff-2016} the cavity state is swapped onto that of an ancilla
qubit. The ancilla is initialized in its
ground state and is dispersively
coupled to the cavity field encoding the parity of the remaining
qubits according to~(\ref{eq:3}). The swapping of the parity
onto the ancilla is achieved in two steps. In the first step, a conditional $\pi$-pulse is
applied to the ancilla qubit
conditioned on the vacuum state of the cavity~\cite{Leghtas-2013a}. This step results in the tripartite entangled state
\begin{align}
\ket{\varphi(T)}=\ket{0}\ket{e}_A\frac{\openone+\bm
  P_N}{2}\ket{\psi_N}+\ket{\beta}\ket{g}_A\frac{\openone-\bm P_N}{2}\ket{\psi_N},
\end{align}
where $\ket{g}_A$ and $\ket{e}_A$ denote the ground and excited states
of the ancilla. In the second step, the cavity is disentangled either via a
conditional displacement of amplitude $-\beta$ conditioned on the ground
state of the ancilla qubit ~\cite{Nigg-2013}, or by
inverting the unitary encoding operations ~\cite{Blumoff-2016}. This results in the
state
\begin{align}
\ket{\varphi(T)}=\ket{0}\left(\ket{e}_A \frac{\openone+\bm
  P_N}{2}\ket{\psi_N}+\ket{g}_A\frac{\openone-\bm P_N}{2}\ket{\psi_N} \right),
\end{align}
where the parity is encoded in the
state of the ancilla. The latter can subsequently be read out via standard
homodyne detection ~\cite{Blais-2004a,Wallraff-2004a}.

An advantage of this readout via an ancilla qubit is that after the
entanglement swapping, the cavity is back in the vacuum state and no
further information about the multi-qubit state can leak out from the
cavity. However, the decoherence of the ancilla does limit the
fidelity of the parity mapping and readout as observed in~\cite{Blumoff-2016}.

\citet{Govia-2015a} proposed an alternative readout based on direct
photon detection via a CBJJ capacitively
coupled to the cavity. The basic idea of this readout, the physical
mechanism of which is explained in details in Section \ref{sec:qualtitative discussion}, is as
follows: In the state of Eq.~(\ref{eq:3}), if a photon is detected, then the
multi-qubit parity is inferred to be even. If a photon is not
detected, then the parity is inferred to be odd with some probability
that depends on the measurement time and the detector efficiency. In
\cite{Govia-2015a} it was shown that this approach leads approximately to a quantum non-demolition parity readout
under the condition that the dispersive shifts of all qubits are
equal. However, for the purpose of stabilizer quantum error correction, QNDness of
parity measurements while necessary is not a sufficient
condition. Indeed the kind of parity measurements required must
preserve the coherence within each parity subspace. This property has
recently been coined eigenstate preserving QND (EP-QND)~\cite{Sarlette-2017}.

One of
the main goals of the present work is to analyze in detail the back-action of the parity measurement based on photon detection~\cite{Govia-2015a}. In Section \ref{sec:qualtitative discussion}, we
show that it is in general not EP-QND because the
emitted photons contain more information than the parity bit alone. To
a lesser extent, this also affects the parity readout used
in~\cite{Nigg-2013, Blumoff-2016}, because the parity encoding and the swapping of the parity
information onto the ancilla take a finite amount of time during
which photons may escape the cavity. In the following we focus on the readout stage of the parity
measurement, once the parity bit has been encoded in a photonic
state such as in Eq.~(\ref{eq:3}).

\section{System and model}
\label{sec:system and model}

\begin{figure}[ht]
\includegraphics[width = 0.4 \textwidth]{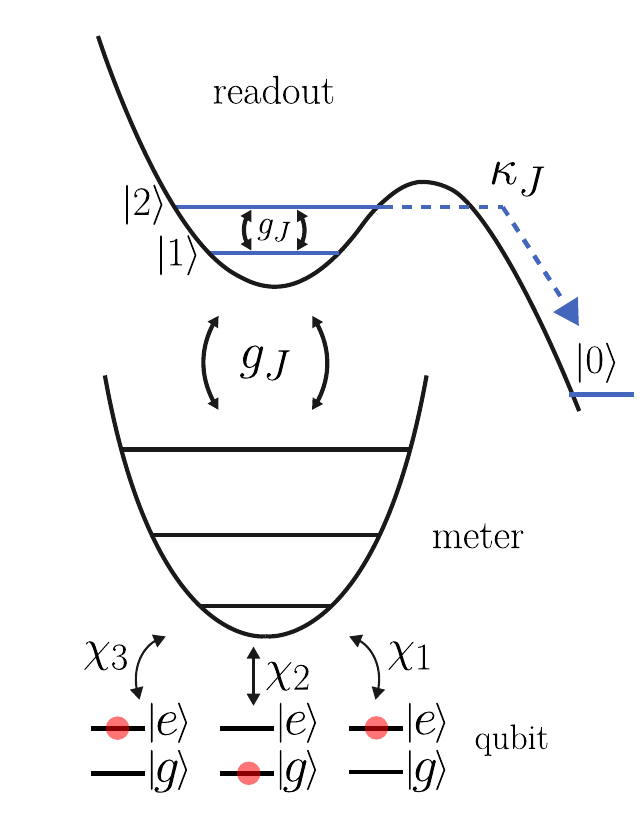}
\caption{(Color online) Model of the parity measurement scheme with
  qubits dispersively coupled to a meter, which in this case is
  modelled by a harmonic oscillator. The qubit parity state can be
  entangled with the meter via a dispersive interaction $\mychi_i$. The
  parity information can be read out via a current biased Josephson junction (CBJJ), which is modelled
  here as a three-level system with states $\ket{0}$, $\ket{1}$ and $\ket{2}$.}
\label{fig:1}
\end{figure}

We consider a specific circuit quantum electrodynamics
(cQED) architecture, where a set of superconducting qubits, such as
e.g. transmon qubits, are dispersively coupled to the quantized
field of a microwave resonator. A concrete realization of this type of
architecture with 3D-transmons and cylindrical microwave resonators is
provided in~\cite{Blumoff-2016} but realizations with coplanar waveguide
resonators are also common~\cite{Riste-2015}. In the following we
consider an abstract model that applies to both implementations. 
The considered system is shown in Fig.~\ref{fig:1} and consists
of a high-Q microwave resonator dispersively coupled to $N$ transmon
qubits and furthermore capacitively coupled to a CBJJ, which serves as
a microwave photon detector~\cite{Romero-2009,Chen-2011,Peropadre-2011,Poudel-2012}. The Hamiltonian we use to model this system is
  
\begin{align}
\begin{split}
\bm H &= \left( \omega_c  + \sum_{i=1}^ N \mychi_i \bm{\sigma}_i^z \right) \bm{a}^\dag \bm{a}+ g_J \left( \bm{a} \ket{2} \bra{1} + \bm{a}^\dag \ket{1} \bra{2} \right) \\
& + \omega_{12}  \ket{2} \bra{ 2} - \omega_{20}  \ket{0}  \bra{0}.
\end{split}
\label{eq:full_syst_Hamiltonian}
\end{align}
Here $\bm\sigma_i^z=\ket{e}\bra{e}_i-\ket{g}\bra{g}_i$ denotes the
Pauli $Z$ operator for qubit $i$. The inevitable
dissipation in the CBJJ associated with the photo-detection process is
accounted for by the Lindblad master equation,
\begin{align}
\bm{\dot{\rho}} = -i \left[ \bm H, \bm{\rho} \right] + \kappa_J \bm {\mathcal{D}} \left[ \ket{0} \bra{2} \right]\bm{\rho},
\label{eq:Lindblad ME}
\end{align}
where $\bm {\mathcal{D}} \left[ \bm c \right]\bm{\rho} = \bm c \bm{\rho} \bm c^{\dag} - \frac{1}{2} \left( \bm c^{\dag } \bm c \bm{\rho} + \bm{\rho} \bm c^{\dag} \bm c \right)$.

Here we have reduced the CBJJ to an effective
three-level system~\cite{Poudel-2012}. The states $\vert 1 \rangle$ and $\vert 2 \rangle$ represent the two states
localized inside a well of the tilted washboard potential of the
CBJJ (see Fig.~\ref{fig:1}). Via the dc current bias, the transition frequency between $\vert 1 \rangle$ and $ \vert 2 \rangle$ is tuned in resonance
with the bare cavity frequency $\omega_c$. Furthermore, by suitably
designing the junction capacitance, it is possible to make the upper
level $\vert 2 \rangle$ couple strongly to the continuum, which is modeled here as an
additional state $\vert 0 \rangle$. A photon leaving the cavity towards the CBJJ
coherently populates level $\vert 2 \rangle$, which incoherently
decays at a rate $\kappa_J$ into the
continuum state $\vert 0 \rangle$. The tunnel coupling of the lower level $\vert 1 \rangle$ to the
continuum state $\vert 0 \rangle$ is exponentially smaller than the coupling between $\vert 2 \rangle$
and $\vert 0 \rangle$ and will be neglected in the following. Note
however that this coupling
will lead to dark counts and thus negatively impact the parity
readout fidelity. For a discussion of this effect see e.g.~\cite{Poudel-2012}.

\section{Qualitative discussion of the measurement back-action}
\label{sec:qualtitative discussion}

\begin{figure}[h!]
\includegraphics[width = 0.50 \textwidth]{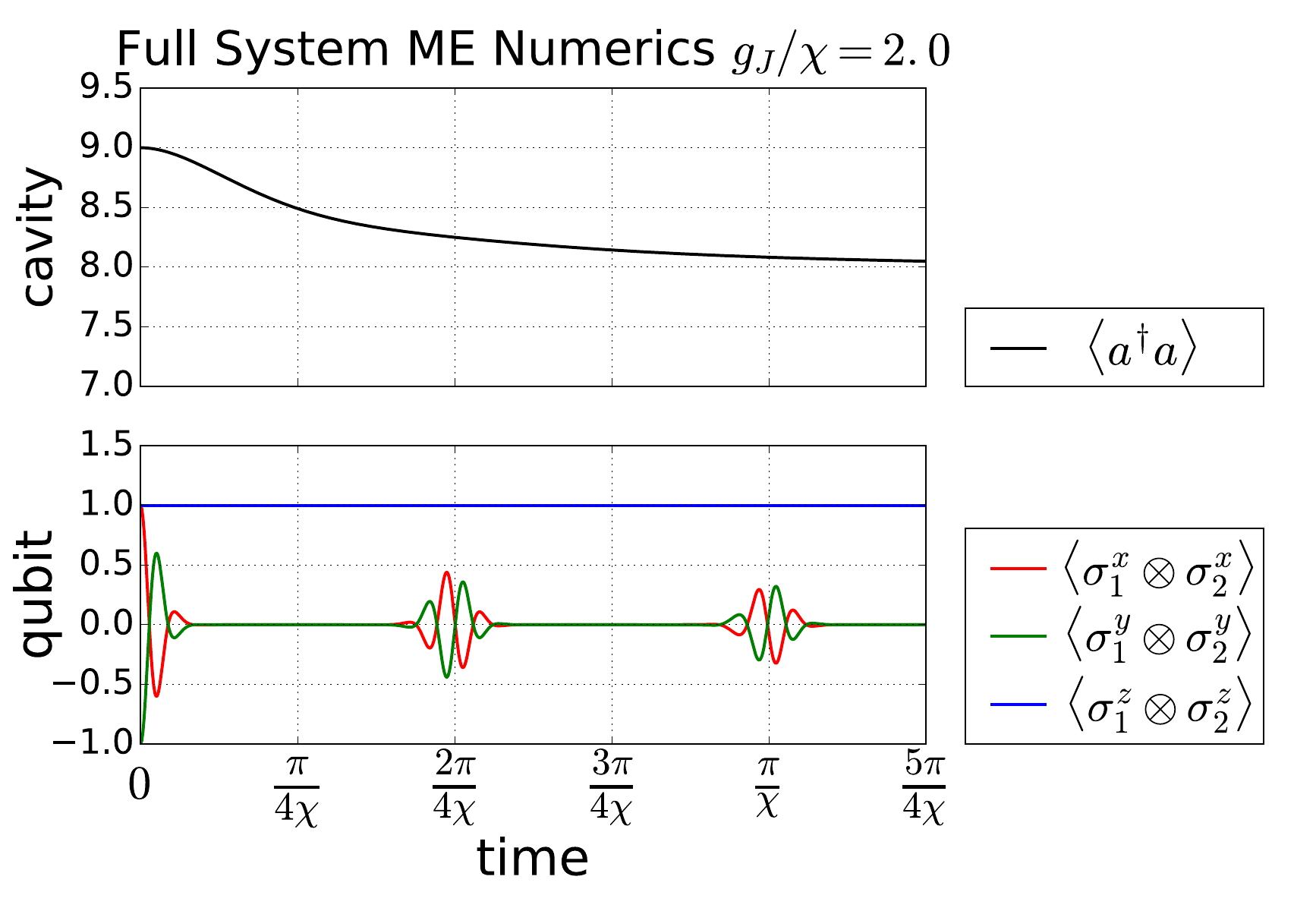}
\caption{Numerical solution of the master equation Eq.~(\ref{eq:Lindblad ME}) with the Hamiltonian Eq.~(\ref{eq:full_syst_Hamiltonian}) for the initial state $\ket{\psi} = 1/ \sqrt{2}(\ket{gg}+ \ket{ee}) \otimes \ket{\alpha} \otimes \ket{1}_{\rm CBJJ}$, with $g_J / \mychi = 2.0$ and $\kappa / g_J = 10.0$. }
\label{fig:full_sys_num}
\end{figure}

In this section we briefly discuss the dynamics of of the full system depicted in Fig.~\ref{fig:1}, obtained by numerically solving the Lindblad master equation Eq.~(\ref{eq:Lindblad ME}).
To illustrate the effect of the measurement we show in Fig.~\ref{fig:full_sys_num} the time evolution of a two-qubit state coupled
to a cavity with amplitude $\alpha$ and the CBJJ. The system is initially in the state $\ket{\psi} = (\ket{gg}+ \ket{ee})/ \sqrt{2} \otimes \ket{\alpha} \otimes \ket{1}_{\rm CBJJ} $. For simplicity we let the dispersive shifts from Eq. (\ref{eq:full_syst_Hamiltonian}) be equal i.e. $\mychi_1 = \mychi_2 = \mychi$.

As a measure for phase coherence within a given parity subspace we use
the expectation values $\braket{\bm\sigma_1^i\bm\sigma_2^i}$ for
$i=x,y$. The hermitian part of the Lindblad master equation
(\ref{eq:Lindblad ME}) leads to a periodical change of
$\braket{\bm\sigma_1^x\bm\sigma_2^x}$ and
$\braket{\bm\sigma_1^y\bm\sigma_2^y}$ which is a consequence of the
entanglement of the qubits with the cavity due to the dispersive
interaction. We will refer to the periodical reappearance of maxima in
these expectation values as the revival of
coherence~\cite{walls2007quantum}. The decrease in the amplitude of
these revivals is a direct measure of the intra parity-subspace
decoherence and is a consequence of the non-hermitian part of Eq.~(\ref{eq:Lindblad ME}), which describes the effect of the measurement when ignoring the measurement record.
In Section \ref{sec:Characterization of measurement back-action} we
will show that at the level of the individual quantum trajectories, the loss of a photon out of the cavity leads to a random phase kick on the qubit parity-subspaces. Because in the master equation one averages over all such random events, this results in the observed suppression of the revival amplitudes.
Furthermore we obtain from the numerics in Fig. \ref{fig:full_sys_num}
that the cavity decay stops after the loss of one photon. This is
because the CBJJ is trapped in the continuum state on a
much longer time scale than that of the actual photon decay.

The revival time scale can be estimated in the coherent limit, i.e. by
considering a reduced system, where the two qubits are coupled to a
cavity, without CBJJ and leakage. The unitary time evolution through
the Hamiltonian $\bm H=\omega_c \bm{a}^\dag \bm{a} +  \mychi
(\bm{\sigma}_1^z + \bm \sigma_2^z)  \bm{a}^\dag \bm{a}$ leads to the
time dependent entanglement of the qubit and the cavity in the form
$\vert \psi (t) \rangle = 1 / \sqrt{2} \left (\ket{gg} \otimes
  \vert \alpha e^{+ 2i \mychi t} \rangle + \ket{ee} \otimes
  \ket{ \alpha e^{-2 i  \mychi t}} \right)$. The revival occurs if the cavity and the qubit are disentangled, i.e. are separable again, 
hence the time of the revival is $t_{\rm rev} = \pi / (2 \mychi)$.\\

\section{Adiabatic elimination of the CBJJ}
\label{sec:Weak coupling limit}

The parity readout discussed in this work takes place in the limit where the effective Rabi coupling between the cavity
and the CBJJ is small compared with the decay rate of the metastable state
of the CBJJ. 
In this regime, the population of the
metastable state of the CBJJ remains small. This allows us to
adiabatically eliminate the CBJJ in the spirit of a Weisskopf-Wigner approximation. In this way we
obtain an analytically tractable and physically transparent
model of the detector where a photon detection
event corresponds simply to a photon loss event out of the
cavity. The rate at which such an event takes place is calculated as
follows.

Consider a system of $N$ qubits in the computational basis state $\ket{j_1,j_2, \cdots , j_N}$ where $j_i \in \{e,g
\}$ are fixed but arbitrary. The associated total state dependent dispersive
shift is $\Delta=\sum_{i=1}^N\sigma_i\mychi_i$, where $\sigma_i=+1$ if
$j_i=e$ and $\sigma_i=-1$ if $j_i=g$. Since the dispersive coupling
commutes with $\bm\sigma_i^z$, the projected Hamiltonian becomes
\begin{align}
\bm H &= \left( \omega_c + \Delta \right) \bm{a}^\dag \bm{a} + g_J \left( \bm{a} \ket{2}\bra{1} + \bm{a}^\dag \ket{1}\bra{2} \right) \nonumber \\
& + \omega_{C} \vert 2 \rangle \langle 2 \vert - \omega_{20} \vert 0 \rangle \langle 0 \vert.
\label{eq:Hamiltonian_for_numerics}
\end{align}
Here we already tuned the CBJJ on resonance with the bare cavity
frequency $\omega_{12} = \omega_c$. To solve the master equation for
this Hamiltonian we notice that the interaction only couples a
closed set of states: $\vert n+1,1 \rangle$, $\vert n, 2 \rangle$ and
$\vert n, 0 \rangle$, where $\vert n \rangle$ is a Fock state with the
photon number $n$ and $\vert 1 \rangle$, $\vert 2 \rangle$ and $\vert
0 \rangle$ represent the states of the CBJJ. If we truncate the
Hamiltonian to this reduced set of basis states and define
\begin{align}
\bm \rho := 
	\begin{pmatrix}
    		\rho_{11} 	&\rho_{12}	 		&\rho_{10} 		\\
    		\rho_{21}  	&\rho_{22}		  	&\rho_{20} 		\\
     	\rho_{01} 	&\rho_{02} 			&\rho_{00} 		\\
 	\end{pmatrix},
\label{eq:rho_as_matrix} 	
\end{align}
where the subscripts $0$, $1$, $2$ are again representing the states
of the CBJJ, the master equation~(\ref{eq:Lindblad ME}) yields a set of coupled linear differential equations,
\begin{subequations}
\begin{align}
\label{DGLs_cavity-JPM1}
& \dot{\rho}_{11} =  -i g_J \sqrt{n+1}(\rho_{21}-\rho_{12}) \\
& \dot{\rho}_{22}= -i g_J \sqrt{n+1}(\rho_{12}-\rho_{21}) - \kappa_J \rho_{22} \\
& \dot{\rho}_{00} = \kappa_J \rho_{22}\label{eq:7} \\
& \dot{\rho}_{12} = -i g_J \sqrt{n+1}(\rho_{22}-\rho_{11}) -  \frac{\kappa_J}{2} \rho_{12} -i \Delta \rho_{12}\\
\label{DGLs_cavity-JPM5}
& \dot{\rho}_{21} = -i g_J \sqrt{n+1}(\rho_{11}-\rho_{22})-\frac{\kappa_J}{2} \rho_{21} +i \Delta \rho_{21}.
\end{align}
\end{subequations}
In a similar manner as in \cite{schondorf2016optimizing} these equations can be solved by Laplace transformation (See Appendix
\ref{Appendix A}) and yield
\begin{align}\label{eq:6}
  \rho_{00}^{(n)}=1-\exp\left\{  -\frac{4tg_J^2(n+1)}{\kappa_J}\left[ 1-\left( \frac{\Delta}{\kappa_J} \right)^2 \right] \right\}.
\end{align}
Here we have added a superscript $(n)$ to emphasize the dependence on
the photon number $n$. The solution for a coherent state $\ket{\alpha}$
in the cavity is obtained by averaging (\ref{eq:6}) over the Poissonian photon number
distribution \cite{govia2016entanglement}. In the large amplitude limit $|\alpha|^2\gg 1$, we can
neglect the relative photon number fluctuations and perform the replacement
$n+1\rightarrow\bar n=|\alpha|^2$. We then obtain
$\rho_{00}\simeq 1-\exp\left( -\kappa_{\rm eff}^{\rm CBJJ}t \right)$,
with
\begin{align}
\kappa_{\rm eff}^{\rm CBJJ} = \frac{4 \Omega_{\bar n}^2}{\kappa_J} \left[ 1 - \mathcal{O}\left(  \frac{\Delta^2}{\kappa_J^2} \right)  \right].
\end{align}
This approximation holds in the limit $\Omega_{\bar n}\equiv g_J \sqrt{\bar n+1}  \ll \kappa_J$ and $\Delta
\ll \Omega_{\bar n}$ where $\bar n=|\alpha|^2$ and $\Omega_{\bar n}$ is
the effective Rabi frequency. The first inequality defines the regime of an
overdamped CBJJ, that directly decays from its excited state $\ket{2}$
to the continuum state $\ket{0}$, without Rabi flopping with the
cavity states $\ket{n+1}$ and $\ket{n}$. The second relation
embodies that the energy is transferred from the cavity to the CBJJ
fast on the time scale characterizing the multi-qubit
dynamics. 
Note
that previous work by \citet{Govia-2014} focused on an intermediate regime where
$\kappa_J\simeq g_J$.

A caveat of the adiabatic elimination is that
we have lost the saturation effect due to the long relaxation time of
the CBJJ (see Section \ref{sec:qualtitative discussion}). This can
however be accounted for a posteriori by matching the effective cavity decay $\braket{\bm
  a^{\dagger}\bm a}=|\alpha|^2e^{-\kappa_{\rm eff}^{\rm cav}t}$ with
the saturation behavior of the CBJJ via $\langle
\bm{a}^\dag \bm{a} \rangle = \lvert \alpha \rvert^2 -
\rho_{00}(t)$. Expanding the population decay of the cavity on the
left hand side and $\rho_{00}(t)$ on the right hand side of this
equation for short
times, we find the effective decay rate
\begin{align}
\kappa_{\rm eff}^{\rm cav} = \frac{4 g_J^2}{\kappa_J} \left(1 - \mathcal{O} \left( \frac{\Delta^2}{\kappa_J^2} \right) \right).
\label{eq:Cavity_decay_rate_detuned}
\end{align}
This form is reminiscent of the resonant vacuum Purcell decay
rate. The second term in the parenthesis represents the effect of the qubit state dependent detuning on the decay
rate. It is of order $\sim\mathcal{O}((\Delta/\kappa_J)^2)$ and is
therefore negligible as long as the relation $\kappa_J \gg \Omega_{\bar n}
\gg \Delta$ holds. In Appendix~\ref{Appendix B}, we discuss the
consequence of this higher-order term on the measurement
back-action. Here we focus on
the leading order measurement back-action, which is independent of the
multi-qubit state and characterized by the effective detection rate
$\kappa_{\rm eff}^{\rm cav}=4g_J^2/\kappa_J$.





\section{Characterization of intra parity-subspace decoherence}
\label{sec:Characterization of measurement back-action}

In the effective model derived in Section~\ref{sec:Weak coupling limit}, where $N$ qubits are coupled to a cavity with
the effective photon detection rate $\kappa_{\rm eff}^{\rm cav}$ the Hamiltonian reduces to 
\begin{align}\label{eq:4}
\bm H=\omega_c \bm{a}^\dag \bm{a} + \sum_i^N \mychi_i \bm{\sigma}_i^z  \bm{a}^\dag \bm{a} 
\end{align}
and the dynamics of the dissipative system can be described by the
Lindblad master equation,
\begin{align}
\label{masterequation}
{ \bm{\dot{\rho}}} = -i \big[ \bm H, {\bm \rho} \big] + \kappa^{\rm cav}_{\rm eff}  \bm D \left[ \bm{a} \right] {\bm \rho}.
\end{align}
However, a master equation is
the average over infinitely many measurements and ignores the outcome
of individual measurements. A clearer picture of the measurement
back-action is obtained from a quantum trajectory analysis which keeps
track of the measurement outcome. Because this measurement is based on
photo-detection, a trajectory consists of a (pure) state conditioned
on the presence or absence of a photon detection event random in
time. The corresponding unraveling of the master equation~(\ref{masterequation}) is
obtained in a standard fashion~\cite{walls2007quantum} by introducing the measurement
operators $\bm M_0= 1- [ iH + \frac{\kappa_{\rm eff}}{2} \bm a^\dag \bm a ]dt$ and $\bm M_1= \sqrt{\kappa_{\rm eff} dt } ~ \bm a$. If no photon is detected in a
given time step the conditional state evolves according to:
\begin{align}
\ket{\psi (t + dt)} = \frac{1}{\sqrt{\bra{\psi} M^\dag_0 M_0 \ket{\psi}}} M_0  \ket{\psi (t)}.
\end{align}
If a photon jump occurs the state evolves according to the jump dynamics
\begin{align}
\ket{\psi (t + dt)} =\frac{1}{\sqrt{\bra{\psi} M^\dag_1 M_1 \ket{\psi}}} M_1  \ket{\psi (t)}.
\end{align}
For parity detection, the initial state is of the form
$\ket{\psi_0}=\ket{\alpha}\bm P_{\rm o}\ket{\psi}_N+\ket{0}\bm
P_{\rm e}\ket{\psi}_N$ (see Eq.~(\ref{eq:3})), where $\ket{\alpha}$ is a
coherent state and $\bm P_{\rm e}=(\openone+\bm P_N)/2$ ($\bm P_{\rm
  o}=(\openone- \bm P_N)/2$) is the projector onto the
even (odd) parity subspace. For compactness we introduce the following
notation for an $N$ qubit basis state
$\ket{\sigma_1,\sigma_2,\dots,\sigma_N}=\ket{(-1)^{n_1},(-1)^{n_2},\dots,(-1)^{n_N}}\equiv
\ket{n}$, where $n$ is the integer with binary representation
$n_1n_2\dots n_N$. With this notation, the multi-qubit state is
$\ket{\psi}_N=\sum_{n=0}^{2^N-1}c_n\ket{n}$ with $\sum_n|c_n|^2=1$ and the parity defined in
Eq.~(\ref{eq:5}) corresponds to the Hamming weight of the binary
representation of the number
$n$. The state right before ($-$) and right after ($+$) a detection
event taking place at time $t_J$ can be
written explicitly as
\begin{align}
\label{eq:stochastic wave function before jump}
\ket{ \psi}_- &= \frac{1}{\mathcal{N}_-}  \sum_{n=0}^{2^N-1}  c_n
                       \bm P_{\rm o}\ket{n}  \ket{ e^{-(i \omega_c  +
                i \Delta_{n} +
                      \frac{\kappa_{\rm eff}}{2})t_J }
                \alpha}
              +\bm P_{\rm e}\ket{\psi}_N\ket{0},\\
\ket{\psi}_+ &=  \frac{1}{\mathcal{N}_+}\sum_{n}   c_n
  e^{ -i (\omega_c +\Delta_{n}) t_J} \bm{P}_{\rm o}\ket{n}\otimes \vert e^{-(i
               \omega_c  +  i \Delta_{n}+\frac{\kappa_{\rm eff}}{2})t_J }  \alpha \rangle .
\end{align}
Here $\Delta_n=\sum_i\mychi_i(-1)^{n_i}$ denotes the total
dispersive shift of the $N$-qubit basis state $\ket{n}$, 
$\mathcal{N}_+= \sqrt{{}_N^{}\!\braket{\psi|\bm P_{\rm o}|\psi}_N}$ and $\mathcal{N}_-= \sqrt{ e^{- |\alpha|^2 (1- e^{- \kappa_{\rm eff} t})} {}_N^{}\!\braket{\psi|\bm P_{\rm o}|\psi}_N + {}_N^{}\!\braket{\psi|\bm P_{\rm e}|\psi}_N}$.


The back-action is now clear. Following a photon loss event, the
state undergoes a phase kick {\em which depends on the
  associated multi-qubit state}, i.e. each component of the
multi-qubit state acquires a {\em different} phase. In addition the amplitude of the
cavity state is exponentially suppressed at the rate $\kappa_{\rm
  eff}^{\rm cav}$. Crucially, because the phase kicks are random, the
dephasing they induce between the multi-qubit components {\em within a
given parity subspace} results, after
averaging, in intra parity-subspace decoherence. The simple physical
picture is that an emitted photon carries more information about the
multi-qubit state than only the parity bit which is encoded in the
presence or absence of a photon. This additional information, which is
encoded in the phase of the emitted photon, is {\em in principle} accessible and
hence its presence must reduce quantum coherence in the same way as for example which-path
information suppresses the ability of a quantum particle
to interfere with itself in a double-slit experiment.

\begin{figure}[h!]
\includegraphics[width = 0.5 \textwidth]{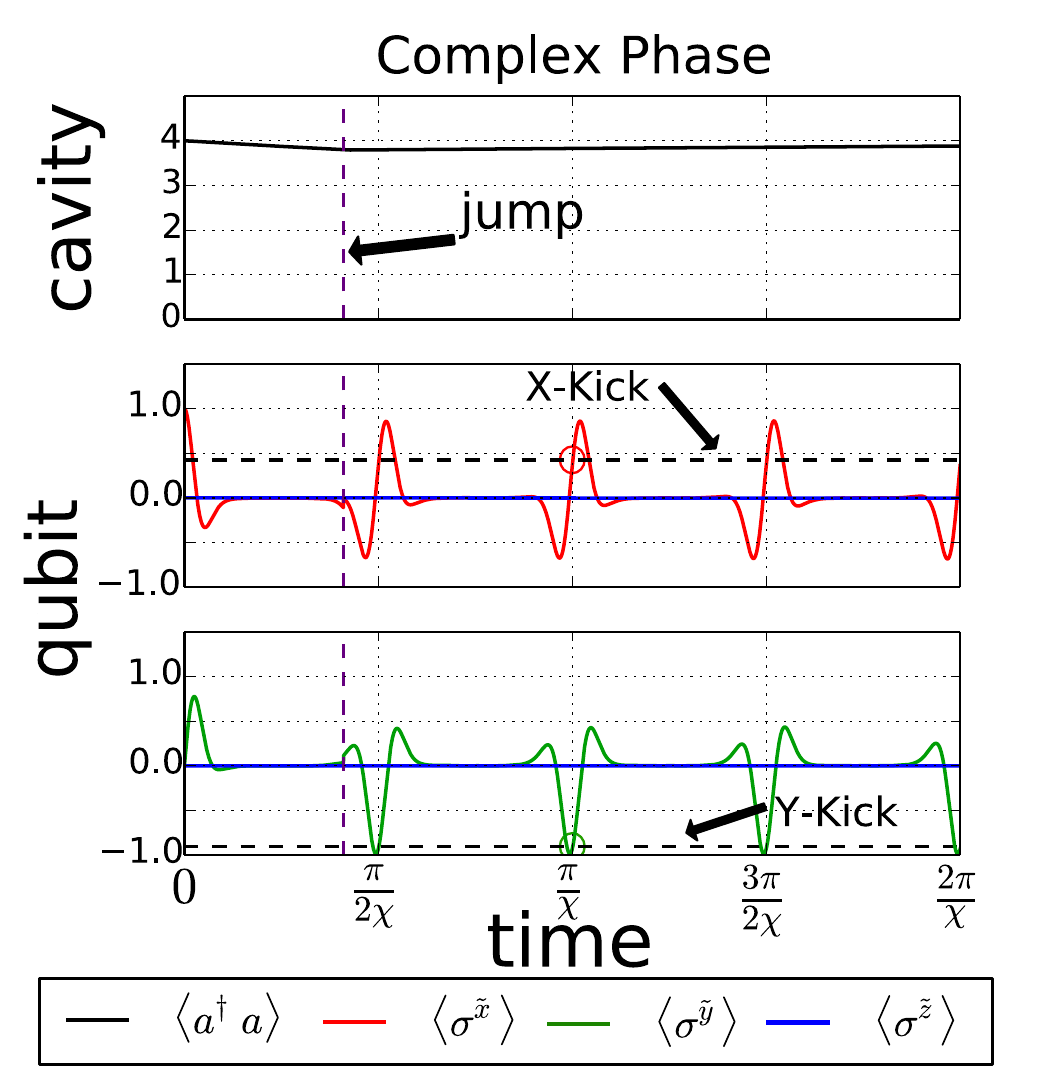}
\caption{(Color online) Comparison between the phase kick prediction of our simple analytic model with the full numeric solution of the system dynamics. The values of the phase kicks predicted by the analytic model (horizontal dashed lines) agree well with the observed kicks at multiple of the revival time $t_{\rm rev}=\pi/2\chi$. The initial state is $1 / \sqrt{2} (\ket{gg} + \ket{ee}) \otimes \ket{\alpha} \otimes \ket{1}_{\rm JPM}$, where $\alpha = 2$ and $\mychi_1 = \mychi_2 = \mychi$. }
\label{fig:phase-kick}
\end{figure}

To confirm this simple interpretation of the dominant source of intra
parity-subspace decoherence, we compare, in Fig.~\ref{fig:phase-kick}, the analytic predictions with a
numerically exact Monte Carlo quantum trajectory simulation of the full system including the
CBJJ dynamics. The system is initialized in the pure state $1 /
\sqrt{2}(\ket{gg}+\ket{ee}) \otimes \ket{\alpha} \otimes \ket{1}$. At
the random jump time $t_J$ a jump occurs (vertical dashed line) at
which the qubit receives a kick. 
The 2-qubit state is initially polarized
in $X$-direction ($\braket{ \sigma_1^{x} \otimes \sigma_s^x} = 1$). 
At the revival, where we can neglect the cavity dynamics the expectation values for the 2-qubit state after the jump are  $\braket{ \sigma_1^{x} \otimes \sigma_2^x} = \cos(2
\mychi t_J)$ and  $\braket{  \sigma_1^{y} \otimes \sigma_2^y} = \sin(2 \mychi
t_J)$. We will refer to this values as X- and Y-Kick. This agrees with the numerical solution obtained in
Fig.~\ref{fig:phase-kick} at the revival times (marked with
dots). We emphasize that in contrast to the master equation result
of Fig.~\ref{fig:full_sys_num},
in the trajectory picture of Fig.~\ref{fig:phase-kick}, the revival
height is not damped since the state remains pure along the trajectory.

Having understood the dominant source of back-action in
photo-detection based parity measurement, we next turn to the question
of how to suppress it. One option would be to use the acquired phase information in a coherent feedback loop to combat decoherence of the multi-qubit
state as shown in~\citet{Kockum-2012}. This should also work in the case where
homodyne detection is used instead of photo-detection via the
CBJJ~\cite{Tornberg-2010,Kockum-2012}. The phase information gathered in this way could then be used
in a coherent feedback loop to combat decoherence of the multi-qubit
state. However, homodyne detection in the weak measurement limit,
would suffer even more from the entangling dynamics due to the
dispersive interaction that is always on. Previous work~\cite{Puri-2016a}
addressed a similar problem by utilizing squeezing to ``hide unwanted
information'' in the enhanced noise of an anti-squeezed
quadrature. Alternatively, the unwanted
entanglement dynamics in the readout phase of the measurement could be
suppressed by using a high-Q tunable resonator, which after the
encoding phase is strongly detuned from the qubits. While progress has recently been achieved with the fabrication of tunable
high-Q microwave resonators~\cite{Vissers-2015,Carvalho-2016}, further
improvements are necessary to make this approach viable. Here we
discuss a simpler and more direct alternative that works for fixed
frequency resonators and uses dynamical
decoupling to minimize the back-action of the CBJJ detector.

\section{Back-action evasion via dynamical decoupling}
\label{sec:back-action evasion via DD}

On the one hand the dispersive interaction of the qubits with the cavity is crucial
for the entanglement of the parity state with the cavity state during the
encoding stage of the measurement. On the other hand it is not desirable
during the readout stage, because it causes the qubit state dependent
detuning $\Delta_n$ and therefore the random phase kicks, which induce decoherence.
Typically, in a high-coherence architecture the dispersive coupling is
not tunable and cannot simply be turned off after the encoding
stage. If high-fidelity single-qubit rotations are available, as is
the case in state-of-the-art superconducting circuits
architectures, we can however effectively cancel the effect of the dispersive
interaction on the system dynamics by periodically flipping all the
qubits on a time scale shorter than the time scale of the entanglement
dynamics $\sim\pi/|\Delta_n|$. 
This can be achieved by repeatedly applying the pattern
\begin{align}
\bm {U X U U X U} \vert \psi \rangle
\end{align}
on the state, where ${\bm U} = \exp \left( -i \bm H \tau \right)$ is
the unitary time evolution operator, $\bm X = \bigotimes_{i=1}^N
\bm{\sigma}_i^x$ is the $N$-qubit flip operator and $2 \tau$ is the time
between two consecutive flips (except the first flip of a measurement,
which is applied after $\tau$).

\begin{figure}[h!]
\includegraphics[width = 0.4 \textwidth]{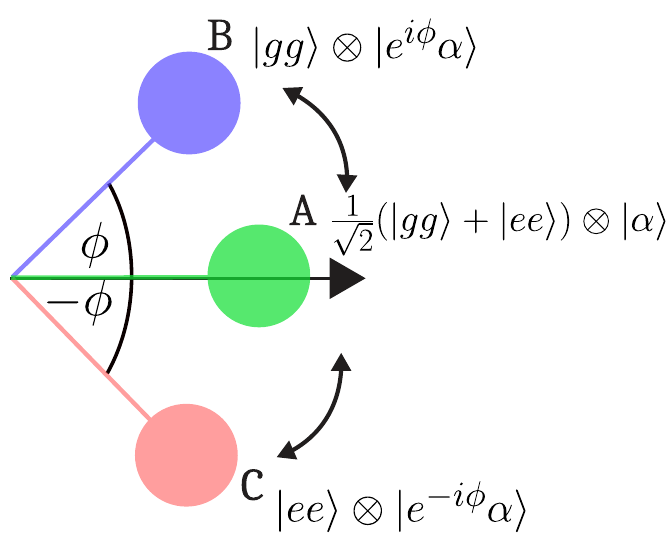}
\caption{(Color online) Cavity phase dynamics of a multi-qubit state
  dispersively coupled to a cavity with qubit flips at the points B and C. The cavity gains different phases 	
depending on the total dispersive shift of the multi-qubit component it is entangled with. E.g. the qubit substates of the initial state $1 / \sqrt{2} (\ket{gg} + \ket{ee}) \otimes \ket{\alpha}$ entangle with cavity states rotating in opposite direction.}
\label{fig:cavity_phase}
\end{figure}

After each flip, the direction of phase rotation of the cavity state,
caused by the dispersive term of the Hamiltonian, is reversed. Figure~\ref{fig:cavity_phase} illustrates the phase dynamics of a multi-qubit state e.g. the state $1 / \sqrt{2} (\ket{gg} + \ket{ee}) \otimes \ket{\alpha}$ in the rotating frame of the bare cavity frequency $\omega_c$ \cite{sete2013catch}. During the time $\tau$ the cavity state entangles with the substates $\ket{gg}$ ($\ket{ee}$) and gains a phase $\phi$ ($- \phi$) according to the time evolution through $\bm U$. It evolves therefore from position A to B (A to C). At this point we flip the qubits by applying the operator $\bm X$, so that during the next unitary time evolution $\bm U$ the cavity state rotates back to its initial position A. Since the qubits are still flipped the cavity will continue to rotate in the same direction during the next time step $\tau$ and the cavity state gains a phase of $- \phi$ ($\phi$) and evolves from A to C (A to B). At this position we apply again $\bm X$ and let it once more evolve according to $\bm U$. This pattern then will be repeated until a photon jump occurs. 
This technique of dynamical decoupling \cite{ernst1990principles} can
be applied on any piecewise constant Hamiltonian $\bm H(t)$, which is
in our case the Hamiltonian of Eq.~(\ref{eq:4}) repeatedly interrupted
by an instantaneous spin flip. If we use the anti-commutation relation
$\{ \bm{\sigma}_i^x , \bm{\sigma}_i^z \} = 0 $ we find that the
sequence $\bm {UXUUXU}$ simplifies to $\exp \left( -4 i  \omega_c \bm
  a^\dag \bm a \tau \right)$, therefore the system will evolve
according to the average Hamiltonian $\bm{\bar{H}} = \omega_c
\bm{a}^\dag \bm{a}$. However this result is only exact, if we can
neglect dissipation. If a photon jump occurs, the assumption of
piecewise constant Hamiltonian does not hold anymore and errors will
be introduced. Furthermore, for simplicity the qubit flips are here assumed to be
instantaneous but more complex sequences of pulses can be designed to
account for finite flip durations~\cite{Khodjasteh-2005}.


\begin{figure}[h!]
\includegraphics[width = 0.50 \textwidth]{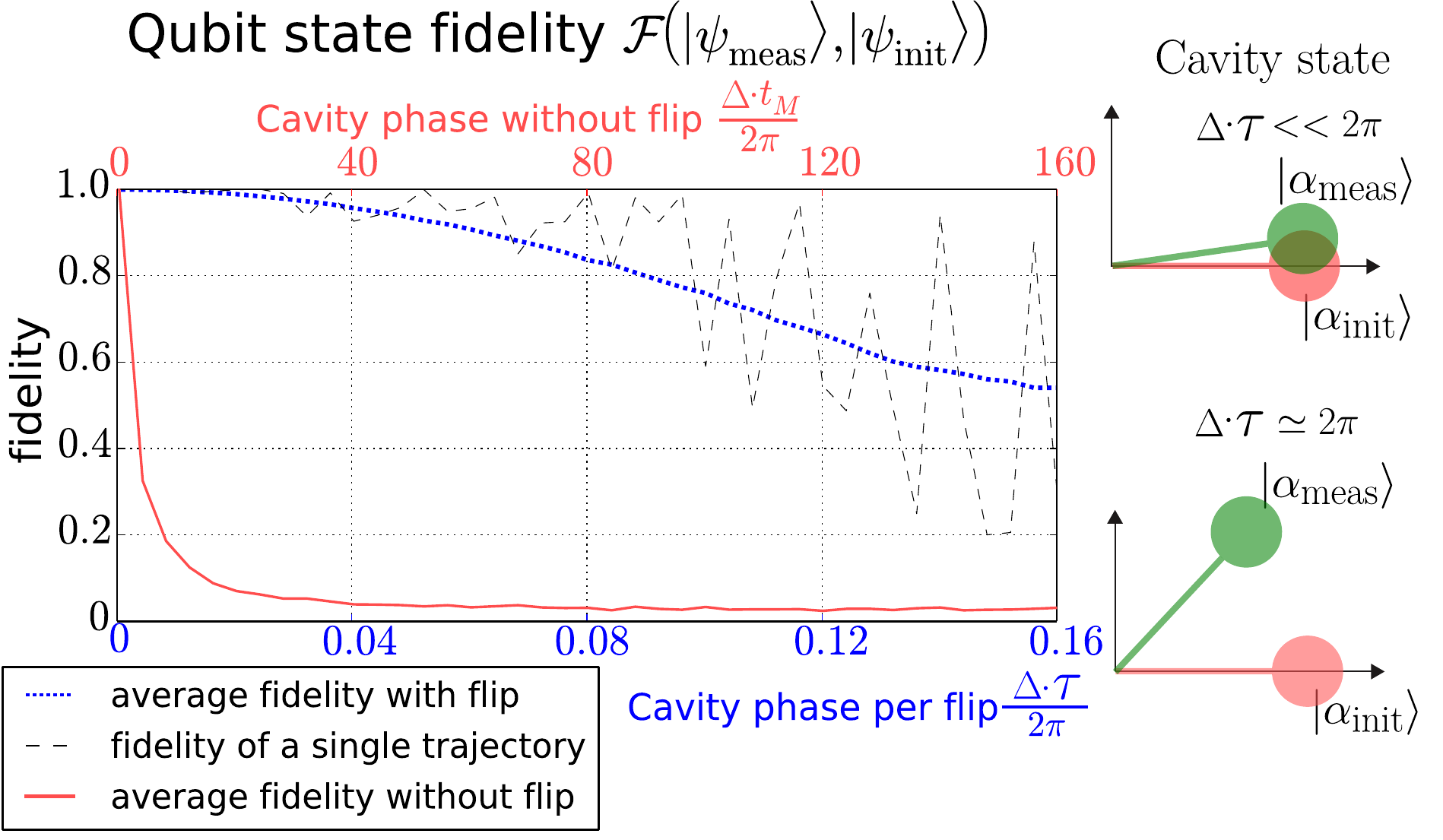}
\caption{(Color online) Average fidelity of the multie-qubit state after
  detection of a photon $\ket{\psi_{\rm meas}}$ with the initial state
  $\ket{\psi_{\rm init}}$ with dynamical decoupling (dotted line) or without dynamical decoupling (solid line).
For a fixed measurement time $t_M$ the fidelity decays fast for
increasing total dispersive shifts $\Delta$ in the non decoupled case. If we apply dynamical
decoupling at a fixed time interval $\tau$ and increase $\Delta$, the
fidelity remains high. The black dashed line represents the fidelity
for a single trajectory for different values of $\Delta$ at a fixed qubit flip rate $\tau$.}
\label{fig:dynamical_decoupling}
\end{figure}

In Figure \ref{fig:dynamical_decoupling} we show that the measurement
fidelity can be high if the phase between the initial
cavity state $\ket{\alpha_{\rm init}}$ and the cavity state at the
photon jump $\ket{\alpha_{\rm meas}}$ is small ($\Delta \tau \ll 2
\pi$). We compare the fidelity of the initial qubit state with the
state after the photon jump for different $\Delta$ at a fixed flip
time interval $\tau$ (dotted line) and for the case where we do not apply
dynamical decoupling for different $\Delta$ at a fixed measurement
time $t_M$ (solid line). Each data point is averaged over $8000$ trajectories. The black
dashed line represents a single trajectory at different $\Delta$ at a
fixed time interval of qubit flips $\tau$ and illustrates the random
character of photodetection for $\Delta \tau \approx 2
\pi$. In this limit of fast cavity rotations the dynamical decoupling breaks
down, if the cavity is far rotated from its initial direction when the
random jump happens. These numerical results provide an upper bound
for the achievable fidelities of about $98\%$. In Table
\ref{Table:realistic values} we estimate achievable fidelities compatible
with state-of-the-art superconducting circuit architectures and the
corresponding qubit flip times $\tau$: The less
phase the cavity gains during a flip, the higher is the fidelity. For
a total dispersive shift of $\Delta=5\,{\rm MHz}$ fidelities above
$90\%$ are reached for switching times on the order of $10\,{\rm ns}$.
\\
\begin{table}[h]
    \begin{tabular}{ | c | c | c | c | c | c |}
    \hline
       $\Delta ~[\rm MHz]$ & $\tau~[\rm ns]$ & phase per flip $\Delta \tau$ &Fidelity $[\%]$ \\ \hline\hline
     10  & 25 &0.04 & 95.5 \\ \hline
     10  & 12.5 &0.02 & 98.8 \\ \hline
     20  & 25 &0.08 & 83.5 \\ \hline
     20  & 12.5 &0.04 & 95.5 \\ \hline
     5 & 25 &0.02 & 98.8 \\ \hline
     5 & 12.5 &0.01 & 99.1 \\ \hline
  \end{tabular}
  \caption{Measurement fidelities of the qubit state for different total dispersive shifts $\Delta$ and time intervals $\tau$. The ratio $\kappa_J / g_J$ is set to $1000$, with $g_J = 10$ MHz and $\kappa_J = 10$ GHz.}
  \label{Table:realistic values}
\end{table}

\section{Detector bias suppression and heralded parity detection}
\label{sec:detector bias}

\begin{figure}[h!]
\includegraphics[width = 0.4 \textwidth]{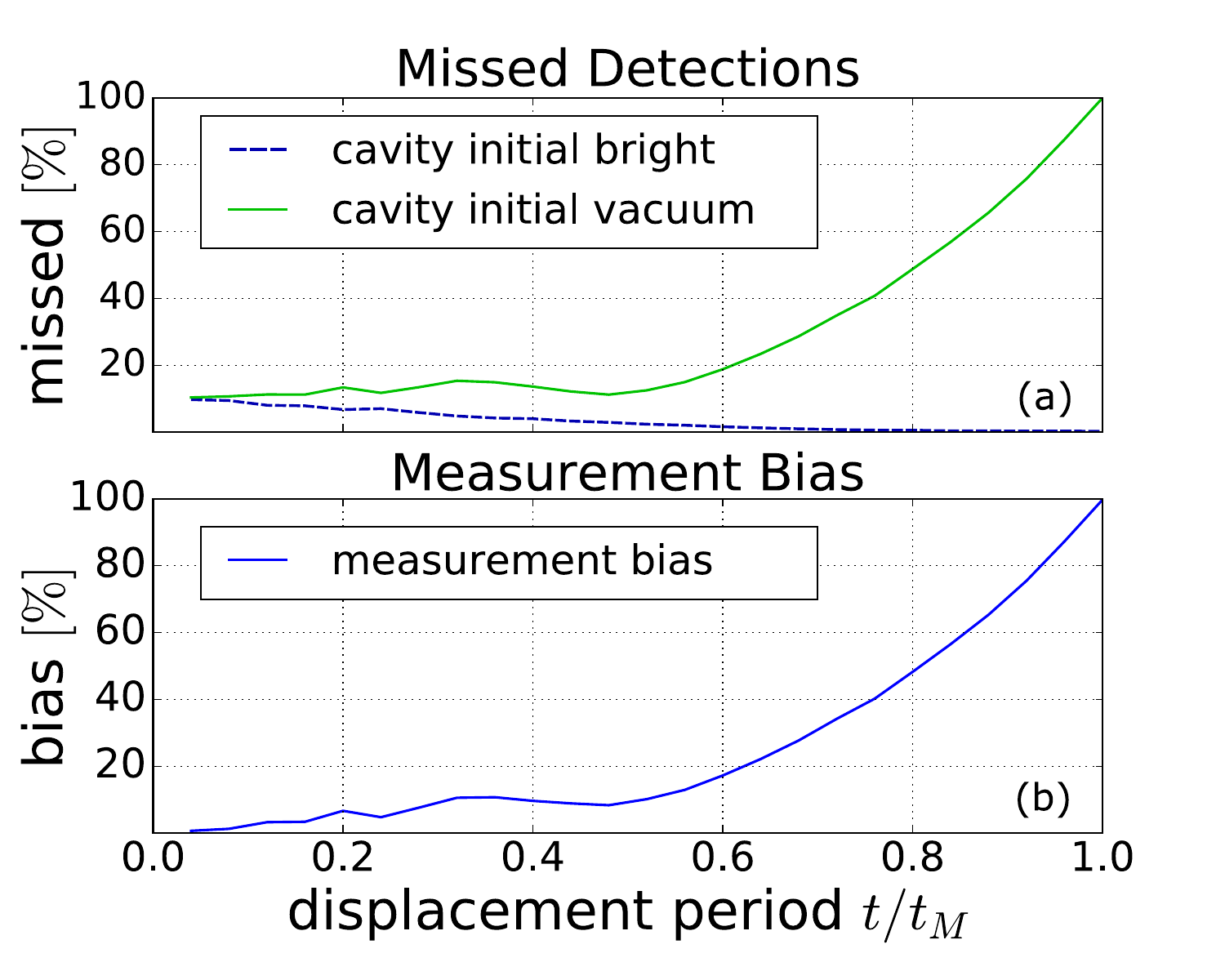}
\caption{(Color online) Missed Detections (a) shows the probability to not detect a photon for an
  initial bright cavity (dashed line) or a cavity initially in the
  vacuum state (solid line) for different cavity displacement periods $t
  / t_M$. The measurement time is set to $t_M = 1 / \kappa_{\rm eff}^{\rm cav}$.
The measurement bias (b) is the difference of missed detections with an initially bright cavity and an initially dark cavity. Since there
cannot be any photon detection in the latter case when $t/t_M=1$, the bias is approximately $ 100 \%$. For a fast displacement the time that the even and the odd qubit parity states are associated with a bright cavity are almost identical and the measurement bias tends to zero. }
\label{fig:Measurement_Bias}
\end{figure}

The multi-qubit parity measurement via direct photon detection has a bias towards one of the parities. Due to finite measurement times $t_M$ the
parity associated with the vacuum cannot be inferred with the same
confidence as the parity associated with the bright cavity state. If we do not detect a photon, there is always a non-zero probability that the cavity is bright and the measurement time was too short to detect a photon decay. If we also include detector efficiencies $\eta < 1$ the measurement bias towards the parity associated with the bright cavity gets even stronger. In order to suppress this bias we apply a sequence of displacement operations to
swap the encoding (even $\leftrightarrow$ bright, odd $\leftrightarrow$ dark) with (even $\leftrightarrow$
dark, odd $\leftrightarrow$ bright) hence "symmetrizing" the roles of
the two parities. Preferably this displacement should be applied only
if a qubit flipping sequence $\bm{U X U U X U}$ is finished, therefore
at integer multiples of $4 \tau$. In this case we know that the cavity
amplitude in the rotating frame of the bare cavity frequency
$\omega_c$ is simply $\lvert \alpha(t)\rvert^2 = \lvert \alpha
\rvert^2 \exp \left( - \kappa_{\rm eff}^{\rm cav} t \right)$. This
procedure will lead to the possibility of a heralded parity
detection: If a photon is detected we know that the qubits are in the
parity state that is associated with the bright cavity state according
to the cavity encoding at the time of detection. If we do not detect a photon during
the measurement time $t_M$ we have to ignore the result, reset and
repeat the measurement. Figure \ref{fig:Measurement_Bias} shows numerical
results averaged over 20000 successful measurement runs for different
displacement periods $t / t_M$. The probability to not measure a
photon (Missed Detections) if the cavity initially is in the vacuum state (solid line)
decreases for faster cavity displacements. For $t / t_M = 1$, if we
do not displace the cavity at all, the probability to not detect a
photon is $100 \%$ because the cavity is dark. Also the probability
to miss a photon if the cavity is initially in a bright cavity state
increases (dashed line). This stems from the fact that an
initial bright cavity does not stay bright for the entire measurement
duration $t_M$ but rather switches between the vacuum and
$\ket{\alpha(t)}$ effectively decreasing the time where one can measure a
photon to $t_M / 2$.
Therefore, for increasing displacement frequencies the measurement
bias is suppressed at the cost of an increasing number of failed
measurements where no photon was detected. The occurrence of the
latter events on the other hand can be reduced by a longer measurement time $t_M$. 

\section{Conclusion}
\label{sec:conclusion}

In conclusion we have characterized an eigenstate preserving
multi-qubit parity measurement scheme based on direct microwave
photo-detection via a current biased Josephson junction. By
dynamically decoupling the dispersive term of the Hamiltonian during
readout qubit decoherence is suppressed. Furthermore, by periodically swapping the
encoding of the parity onto bright and dark states the measurement
bias can be reduced. The detection of a photon then heralds a
successful parity measurement. We estimated numerically that high
fidelities can be obtained with switching rates on the order of
$100\,{\rm MHz}$ for dispersive couplings of the order of $5\,{\rm
  MHz}$. Finally, we note that although we focused here on a simple
microwave photon detector, the CBJJ, the presented parity measurement
also works with more sophisticated detectors such
as~\cite{Kyriienko-2016, Inomata2016}.



{\it Acknowledgments:} 
This work was financially supported by the Swiss National Science
Foundation and the Spanish MINECO (FOQUS FIS2013-
46768, QIBEQI FIS2016-80773-P and Severo Ochoa Grant
No. SEV-2015-0522), Fundaci\'{o}n Cellex, and Generalitat de Catalunya (Grant No. SGR 874 and 875, and CERCA
Programme). 
P.H also acknowledges funding from the European Union’s Horizon 2020 research and innovation programme under the Marie Sk\l{}odowska-Curie grant agreement No 665884.
We thank Christoph Bruder for useful discussions and the
anonymous referee for suggesting to us the idea of using a tunable
high-Q resonator to mitigate the impact of measurement back-action. The numerical
computations were performed in a
parallel computing environment at sciCORE (\url{http://scicore.unibas.ch/}) scientific computing core facility
at University of Basel using the Python library QuTip (\url{http://qutip.org/}).

\begin{appendix}

\section{Laplace transformation and derivation of $\rho_{00}^{(n)}$}
\label{Appendix A}

To derive the effective decay rate caused by the cavity-CBJJ interaction we describe the system with the Hamiltonian Eq.~(\ref{eq:Hamiltonian_for_numerics}) and solve the master equation (\ref{eq:Lindblad ME}). We can rewrite the Hamiltonian in the reduced set of basis sates $ \vert n+1 , 1 \rangle , \vert n,2 \rangle , \vert n,0 \rangle $ as
\begin{align}
H_n = 
	\begin{pmatrix}
    		(n+1) (\omega_c + \Delta) 	&\sqrt{n+1}g_J  		&0 			\\
    		\sqrt{n+1}g_J  	&(n+1) \omega_c + n \Delta 	&0 			\\
     	0 				&0 					&- \omega_0 	\\
 	\end{pmatrix}
\end{align}
The master equation~(\ref{eq:rho_as_matrix}) then yields the
differential equations (\ref{DGLs_cavity-JPM1} -
\ref{DGLs_cavity-JPM5}). To simplify the calculation we split $\rho_{12}$ into its imaginary and real parts
\begin{align}
& \dot{\rho}_{12}^R =\frac{1}{2} (\rho_{12} + \rho_{21}) = -\frac{\kappa_J}{2} \rho_{12}^R + \Delta \rho_{12}^I\\
& \dot{\rho}_{12}^I =\frac{1}{2 i} (\rho_{12}- \rho_{21}) = - g_J \sqrt{n+1}(\rho_{22}-\rho_{11})-\frac{\kappa_J}{2} \rho_{12}^I - \Delta \rho_{12}^R.
\end{align}
We next apply a Laplace transformation, with $\rho_{11} (0) = 1$ and
find
\begin{align}
\label{eq:rho12R}
& \rho_{12}^R = \frac{\Delta }{s + \frac{\kappa}{2} } \rho_{12}^I \\
\label{eq:rho22}
& \rho_{22} = \frac{2 \Omega_n}{s + \kappa} \rho_{12}^I = \frac{2 \Omega_n}{s + \kappa} \rho_{12}^I\\
& \rho_{11} = \frac{1}{s} - \frac{2 \Omega_n}{s } \rho_{12}^I\\
\label{eq:rho12I}
& \rho_{12}^I = \frac{\Omega_n}{s( \frac{\Delta^2}{s + \frac{\kappa}{2}} +2 \Omega_n^2 (\frac{1}{s}+\frac{1}{s+\kappa}))}
\end{align}
Here we used the shorthand notation $\Omega_n=g_J \sqrt{n+1}$. We are
interested in $\rho_{00}$ which can be obtained by integrating
$\rho_{22}$ (See Eq.~(\ref{eq:7})). To find $\rho_{22}$ we substitute $\rho_{12}^I$ into
Eq.~(\ref{eq:rho22}). The inverse Laplace transform is obtained from the integral:
\begin{align}
\rho_{kl}(t) = \frac{1}{2 \pi i} \int_{\gamma - i \infty}^{\gamma + i \infty} ds~ \rho_{kl}(s) ~e^{st},
\end{align}
which can easily be solved by summing over the residua of the integrand. The poles of $\rho_{12}^I$ are:

\begin{widetext}

\begin{align}
& s_{0} = \frac{1}{2} \left( + \frac{\sqrt{+ \sqrt{ \left(16 \Omega_n^2+\kappa ^2+4 \Delta ^2\right)^2-64 \Omega_n^2 \kappa ^2 }-16 \Omega_n^2+\kappa ^2-4 \Delta ^2}}{\sqrt{2}}-\kappa \right) \\
& s_{1} = \frac{1}{2} \left( - \frac{\sqrt{- \sqrt{ \left(16 \Omega_n^2+\kappa ^2+4 \Delta ^2\right)^2-64 \Omega_n^2 \kappa ^2 }-16 \Omega_n^2+\kappa ^2-4 \Delta ^2}}{\sqrt{2}}-\kappa \right) \\
& s_{2} = \frac{1}{2} \left( + \frac{\sqrt{- \sqrt{ \left(16 \Omega_n^2+\kappa ^2+4 \Delta ^2\right)^2-64 \Omega_n^2 \kappa ^2 }-16 \Omega_n^2+\kappa ^2-4 \Delta ^2}}{\sqrt{2}}-\kappa \right) \\
& s_{3} = \frac{1}{2} \left( - \frac{\sqrt{+ \sqrt{ \left(16 \Omega_n^2+\kappa ^2+4 \Delta ^2\right)^2-64 \Omega_n^2 \kappa ^2 }-16 \Omega_n^2+\kappa ^2-4 \Delta ^2}}{\sqrt{2}}-\kappa \right).
\end{align}

\end{widetext}
The Laplace transfom of $\rho_{22}$ then takes the form
\begin{align}
\rho_{22}(s)=\frac{2 \Omega_n  (s+ \frac{ \kappa}{2} ) (s+\kappa)}{(s-s_0)(s-s_1)(s-s_2)(s-s_2)},
\end{align}
and the inverse transform is $\rho_{22} (t) = \sum_{s_i} {\rm
  Res}(\rho_{22}(s) e^{st}, s_i)$. Finally the occupation of the
continuum state is obtained from $\rho_{00}(t)=\kappa_J \int_0^t \rho_{22}(\tau) d \tau$.
By inspection of the residua we see that ${\rm Res}(\rho_{22}, s_3)$
is the dominant contribution to $\rho_{22}$ which simplifies the
expression for $\rho_{00}$ to
\begin{widetext}
\begin{align}
{\rho}_{00} &= 1-  \exp \left[ \frac{t}{4}  \left( \sqrt{-32 \Omega_n ^2+ 2 \sqrt{256 \Omega_n ^4+32 \Omega_n ^2 \left(4 \Delta ^2-\kappa_J ^2\right)+\left(\kappa_J ^2+4 \Delta ^2\right)^2}+2 \kappa_J ^2- 8 \Delta ^2}-2 \kappa_J \right) \right] \approx 1- \exp \left(- \frac{4 \Omega_n^2}{\kappa} t \left( 1 - 4 \frac{\Delta^2}{\kappa_J^2} \right) \right),
\end{align}
\end{widetext}
where we made use of the limits $ \Delta \ll \Omega_n \ll \kappa_J$.




\section{Higher-order decoherence}
\label{Appendix B}

In Section \ref{sec:Weak coupling limit} we derived the detuning
dependence of the effective decay rate, with a fixed detuning
$\Delta$. Because the detuning $\Delta = \sum_{i}\mychi_i\sigma_i$
depends on the multi-qubit state this means that the effective decay
rate can be different for different multi-qubit state components even
{\em within} a given parity subspace. As a consequence, in addition to
random phase kicks that correspond to amplitude preserving random rotations of the
multi-qubit state around the logical $Z$ axis, these higher-order terms
will lead to random rotations out of the logical $XY$ plane. Because
the detuning dependence is quadratic and for two qubits $\Delta_{\rm
  even}=-\Delta_{\rm odd}$ we must consider at least three qubits to
observe this higher-order effect.

\begin{figure}[h!]
\includegraphics[width = 0.51 \textwidth]{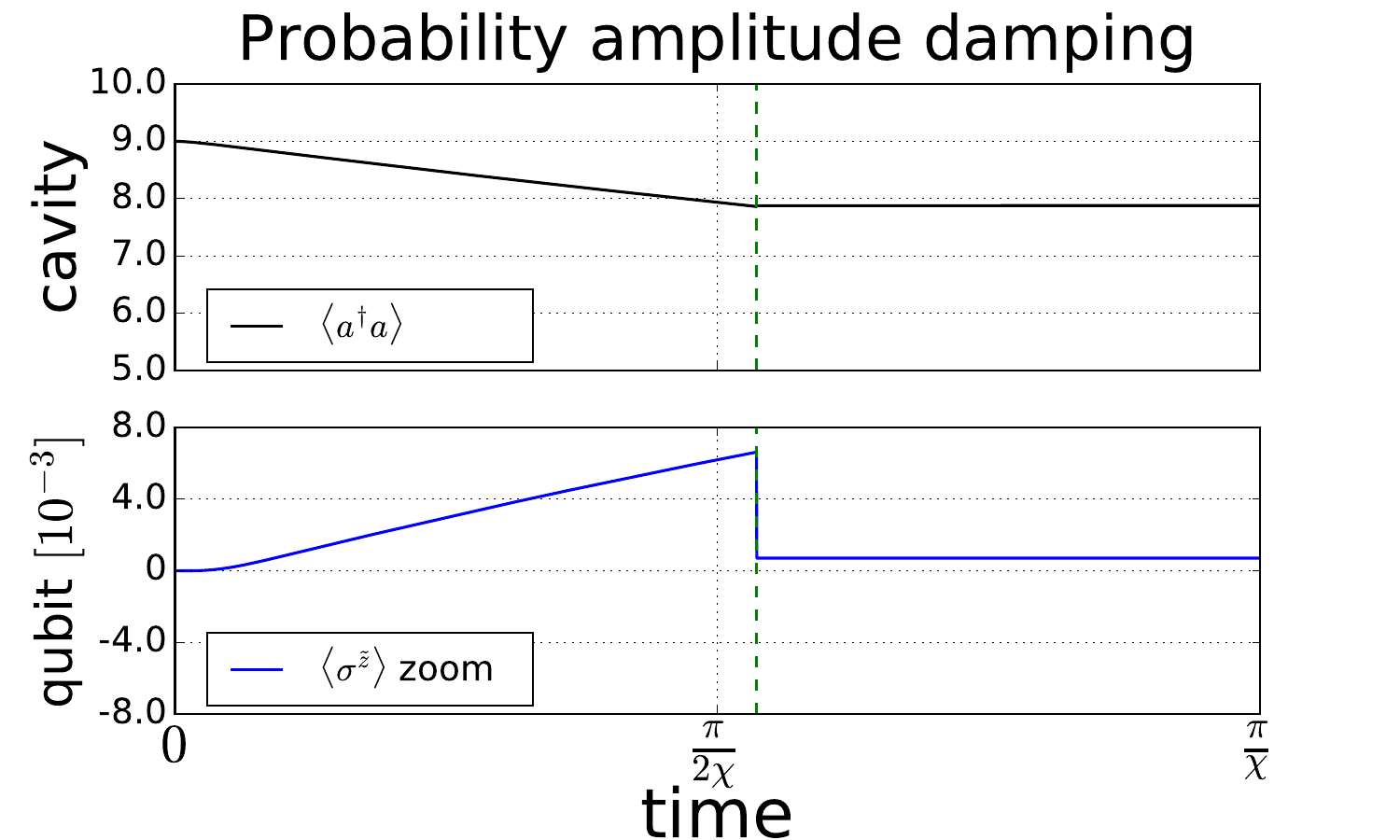}
\caption{(Color online) A full system numerical solution of the quantum trajectory to visualize higher-order decoherence caused by the qubit-state dependent effective decay rate. This numerical result was obtained for $\Omega_n / \mychi = 10$ and $\kappa_J / \Omega_n = 15$. The vertical dashed line indicates the jump time, $\sigma^{\tilde{z}}$-zoom shows the real shift in the probability amplitudes. }
\label{fig:Amplitude damping}
\end{figure}

In Fig. \ref{fig:Amplitude damping} we numerically solve for a quantum trajectory
of the full system with the Hamiltonian from
Eq.~(\ref{eq:full_syst_Hamiltonian}) for the odd three-qubit state
$\ket{\psi} = 1 / \sqrt{2}(\ket{egg} + \ket{eee})$ coupled to a cavity
with $\alpha = 3$ and the CBJJ initially in the state $\ket{1}$. We
define the logical Z-operator $\bm \sigma^{\tilde{z}}= \ket{egg} \bra{egg}-
\ket{eee}\bra{eee}$. Its expectation displays a clear jump when a
photon loss event occurs. Upon averaging over many
trajectories this results in an additional contribution to the
measurement induced decoherence rate. It remains an open question how
to extend the dynamical decoupling scheme to compensate also for such
higher-order effects.

\end{appendix}
\bibliographystyle{apsrev}
\bibliography{Refs}

\end{document}